\documentclass[11pt]{article}

\usepackage{url}
\usepackage{enumerate,mathrsfs,comment}
\usepackage[numbers]{natbib}
\usepackage[OT1]{fontenc}
\usepackage[hypertexnames=false, colorlinks,linkcolor=red,anchorcolor=blue,citecolor=blue]{hyperref}
\usepackage{fullpage}

\usepackage[protrusion=false,expansion=true]{microtype}

\RequirePackage{amsmath}
\RequirePackage{amssymb}
\RequirePackage{amsthm}
\RequirePackage{bm}
\RequirePackage{url}
\usepackage{natbib}
\usepackage{multirow}
\usepackage{graphicx}
\usepackage{subfigure}
\usepackage{makecell}
\usepackage{booktabs}
\usepackage{array}
\usepackage{url}
\usepackage{algorithm}
\usepackage{algorithmic}
\usepackage{dsfont}

\let\hat\widehat

\newtheorem{proposition}{\indent \bf Proposition}
\numberwithin{equation}{section}

\theoremstyle{plain}
\numberwithin{equation}{section}

\newtheorem{lemma}{\indent \bf Lemma}

\newtheoremstyle{mytheoremstyle} %
    {\topsep}                    %
    {\topsep}                    %
    {\normalfont}                   %
    {}                           %
    {\bfseries}                   %
    {.}                          %
    {.5em}                       %
    {}  %

\theoremstyle{mytheoremstyle}

\ifx\BlackBox\undefined
\newcommand{\BlackBox}{\rule{1.5ex}{1.5ex}}  %
\fi

\ifx\QED\undefined
\def\QED{~\rule[-1pt]{5pt}{5pt}\par\medskip}
\fi

\ifx\proof\undefined
\newenvironment{proof}{\par\noindent{\bf Proof\ }}{\hfill\BlackBox\\[2mm]}
\fi

\ifx\theorem\undefined
\newtheorem{theorem}{Theorem}
\fi
\ifx\example\undefined
\newtheorem{example}[theorem]{Example}
\fi
\ifx\property\undefined
\newtheorem{property}{Property}
\fi
\ifx\lemma\undefined
\newtheorem{lemma}[theorem]{Lemma}
\fi
\ifx\proposition\undefined
\newtheorem{proposition}[theorem]{Proposition}
\fi
\ifx\remark\undefined
\newtheorem{remark}[theorem]{Remark}
\fi
\ifx\corollary\undefined
\newtheorem{corollary}[theorem]{Corollary}
\fi
\ifx\definition\undefined
\newtheorem{definition}[theorem]{Definition}
\fi
\ifx\conjecture\undefined
\newtheorem{conjecture}[theorem]{Conjecture}
\fi
\ifx\fact\undefined
\newtheorem{fact}[theorem]{Fact}
\fi
\ifx\claim\undefined
\newtheorem{claim}[theorem]{Claim}
\fi
\ifx\assumption\undefined

\fi
\numberwithin{equation}{section}
\numberwithin{theorem}{section}
\usepackage{multirow}

\long\def\comment#1{}

\begin{document}
\title{DVE: Dynamic Variational Embeddings with Applications in Recommender Systems}

\author
{
    Meimei Liu \quad
    Hongxia Yang%
}

\date{}
\maketitle

\begin{abstract}
 Embedding is a useful technique to project a high-dimensional feature into a low-dimensional space, and it has many successful applications including link prediction, node classification and natural language processing. Current approaches mainly focus on static data, which usually lead to unsatisfactory performance in applications involving large changes over time. How to dynamically characterize the variation of the embedded features is still largely unexplored.  
  In this paper, we introduce a dynamic variational embedding (DVE) approach for sequence-aware data based on recent advances in recurrent neural networks. DVE can model the node's intrinsic nature and temporal variation explicitly and simultaneously, which are crucial for exploration. We further apply DVE to sequence-aware recommender systems, and develop an end-to-end neural architecture for link prediction.  
\end{abstract}


\section{Introduction}
 Graph embeddings aim to learn a low-dimensional representation for each node in a graph accurately capturing relationships among the nodes. This has wide applicability in many graph analysis tasks including node classification \cite{bhagat2011node}, clustering \cite{ding2001min}, recommendation \cite{liben2007link}, and visualization \cite{maaten2008visualizing}. Various embedding methods have been proposed, including classical spectral embedding algorithms (\cite{balasubramanian2002isomap}, \cite{de2002locally}, \cite{belkin2002laplacian}), factorization (\cite{ahmed2013distributed}), neural embedding (\cite{he2017neural}, \cite{hamilton2017inductive}).

\begin{figure}[h]
\centering
  \begin{tabular}{cc}
    \includegraphics[scale=0.4]{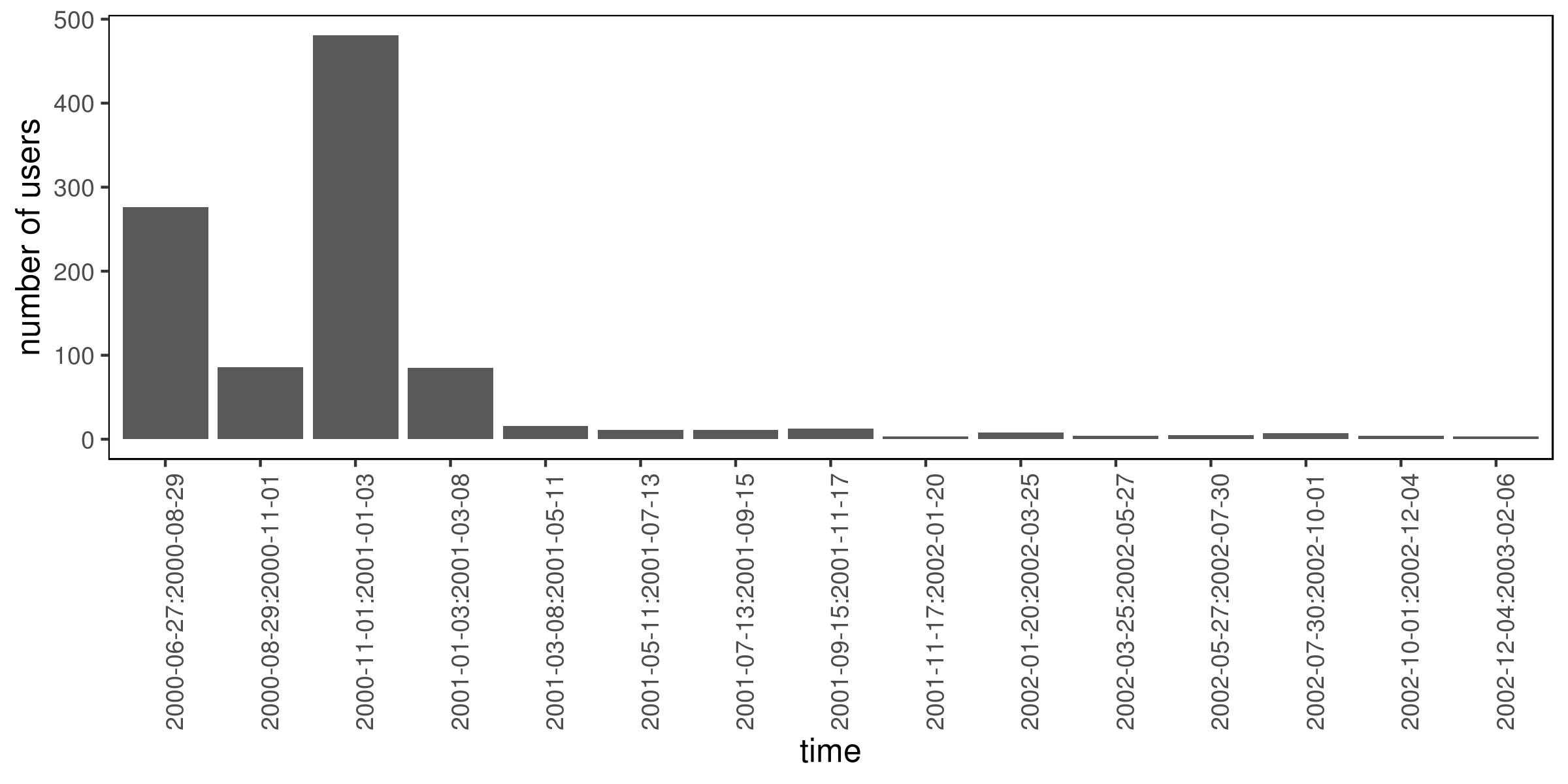} &  \includegraphics[scale=0.4]{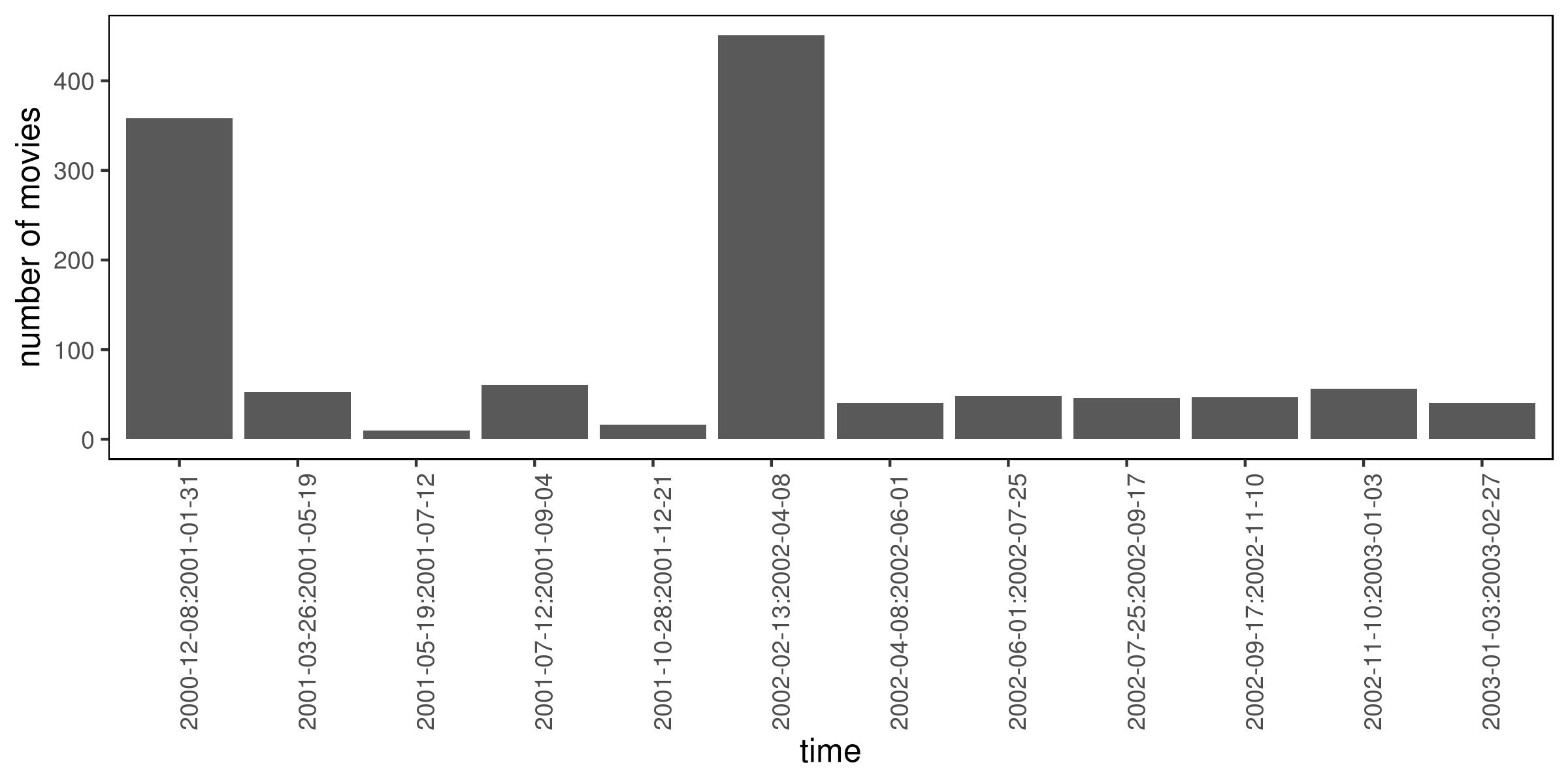}\\
    ~~~~~$(a)$ &  ~~~~~ $(b)$\\
     \end{tabular}
     \caption{ $(a)$: the barplot shows the number of ratings versus time for movie ``The Perfect Storm'' which is released on June, 2000. $(b)$: the barplot shows the number of ratings versus time for the user with user\_id 423.}\label{fig:movielens}
\end{figure}

The above approaches aim to provide point estimators of embedding features, and a significant shortcoming is their inability to express variation, especially the dynamic variation when data are sequentially collected. Being able to accurately represent variation is critical. In sequence-aware recommender systems (RS), where data are collected from sessions or transactions, besides the intrinsic features, user preferences for products varies over time. Item popularity also changes with time. We use the user-movie rating in the benchmark dataset MovieLens (\cite{harper2016movielens}) as an illustration example. Figure \ref{fig:movielens} (a) displays that for a particular movie, the number of ratings varies with time; Figure \ref{fig:movielens} (b) shows that for a certain user, the number of ratings also varies with time. This example reveals the users' behavior and item popularity may dynamically change based on factors such as the available time, interest shift, and environment changes. These dynamic changes suggest not relying too much on exploitation of past behavior in favor of exploration. However, research focusing on this aspect is still lacking; see Section \ref{sec:related_ref} for detailed reference review. 

In this paper, we propose a novel dynamic variational embedding (DVE) method for sequence-aware data to address the aforementioned gap in quantifying variation. We assume that the embedding feature for each node consists of two parts: one is the intrinsic nature and one is the variational feature that captures temporal changes sequentially. 
Recurrent neural networks (RNN) have been shown to be successful at capturing the nuances of nodes' interactions in the short and long term (\cite{sutskever2014sequence}, \cite{lecun2015deep}). Therefore, we develop an RNN architecture to characterize the dynamic variance. We design an embedding feature that can characterize both short and long range dependence. A distinguishing feature of our method is that the individualized dynamic variance can be explicitly included in the model, providing a strong guidance in exploration.  

Applying DVE to sequence-aware recommender systems (RS), we further develop an end-to-end deep neural network (DNN) to study link prediction. Given the explosive growth of information available on the web, RS have been widely adopted by many online services, including e-commerce, and social media sites. Personalized RS is an essential demand for facilitating a better user experience. One of the most popular RS approaches is collaborative filtering (CF) (\cite{sarwar2001item}, \cite{schafer2007collaborative}) that aims to model the users' preference on items based on their previous behavior (e.g., ratings, clicks, buy). 
Among the various CF techniques, a mainstream is measuring the interactions between users and items through products of their latent features (\cite{koren2008factorization}). However, it has been shown in \cite{he2017neural} that such inner product-based models may not be sufficient to capture the complex structure of user interaction data. 

DNN is flourishing in recent years (\cite{sutskever2014sequence}, \cite{he2017neural} etc). It endows the model with a large level of flexibility and non-linearity to learn the interactions between the embedding features of users and items. In this paper, we built a sequence-aware RS by fully utilizing a neural collaborative filtering framework based on DVE. The main contributions of our work are summarized below.

\begin{itemize}
\item We propose a novel dynamic variational embedding (DVE) approach to learn nodes' intrinsic and variational features simultaneously. The dynamic variational feature is achieved by introducing a recurrent neural network (RNN) into the neural embedding architecture. This is crucial for facilitating exploration.
\item We consider a sequence-aware recommender system, and show that handling temporal information plays a vital role in 
improving the accuracy of the RS. 
\item Based on DVE, we develop an end-to-end deep neural architecture for our sequence-aware recommender system, where user's and item's embedding features exhibit temporal dependencies, to study the link prediction. The whole neural architecture is constructed in two parts: one is the embedding layers  for DVE, and one is the neural collaborative filtering layers to explore the non-linear interaction between users and items. 
\end{itemize}

\vspace{-5pt}

\section{Related Works}\label{sec:related_ref}
One classical direction of embedding is factorization-based; examples include spectral embedding algorithms like IsoMap, LLE, Laplacian eigenmap in \cite{balasubramanian2002isomap}, \cite{de2002locally}, \cite{belkin2002laplacian}, and matrix factorization in (\cite{ahmed2013distributed}). 
Neural networks are also used in graph embedding in recent years. 
Deep neural networks have proven successful due in part to their ability to model complicated non-linear data representations. The neural collaborative filtering (NCF) proposed by \cite{he2017neural} fuses matrix factorization and one-hot embedding, and feeds them into a deep neural network framework, showing significant gains in accuracy in prediction. Recently, Graphsage \cite{hamilton2017inductive} proposes an inductive learning approach for node features based on graph convolutions, having wide applicability in massive graph problems. However, the above methods aim to provide single estimator, without characterizing variation. Latent space embedding is one approach to learn features with variation, i.e., using latent representations to characterize features of each node with Bayesian probabilistic models, including latent space models \cite{hoff2002latent}. 

A variety of sequence-aware recommender systems have been proposed in the literature. \cite{koren2009collaborative} developed a time-aware factor model to address the temporal changes in collaborative filtering. \cite{ying2018sequential} developed attention-based RS based on DNN. Again, these methods are lacking variation quantification compared with our DVE based recommender architecture. 

Variational autoencoders (VAEs), combining the deep latent variable model and variational learning technique, are popular in the application of recommender systems recently.  \cite{li2017collaborative} proposes collaborative variational autoencoder (CVAE) approach to learn the item-based embedding in an unsupervised manner. \cite{liang2018variational} constructs a generative model with multinomial likelihood for each user’s preference on all items by assigning a low dimensional latent vector for the user's preference. 

The above approaches are either item-based or user-based unsupervised learning; while our approach is supervised learning, and can learn the variational features of the users and items simultaneously. Another crucial limitation of current VAE-based recommender learnings is their insufficiency in exploration, since the key idea of VAE (e.g.,\cite{liang2018variational}) is minimizing the KL distance between the input behavior and its latent representer, which only focuses on the exploitation of the previous behaviors. However, our DVE-based approach considers both the long-term feature and the dynamic variation for each user and item, thus enabling exploration. Furthermore, the computational and storage bottleneck of the above user-based VAEs become critical for RS with millions of users and items, since the input and the training target include the whole preference for each user. In contrast, the input of our approach includes only the individualized records at time $t$, such as (user\_id, item\_id, click (or score)) at $t$, i.e., the nonzero entries in the sparse preference matrix. Thus we enjoy relatively high computational efficiency. 

Our work is related to the study of NCF in \cite{he2017neural}, but
distinguished from \cite{he2017neural} by the following aspects: (1) our embedding feature is random and dynamic by considering the temporal information; (2) we consider sequence-aware RS instead of static RS.

\section{Method}
The key idea behind our dynamic variational embedding (DVE) approach is that we assume the embedding features have both an intrinsic and variational nature. In the following, we first introduce neural variational embedding algorithm in which the embedding is learned in two parts: the (intrinsic) mean and its variance. Based on such structure, we further illustrate the construction of DVE, incorporating temporal changes into the variance by employing recurrent neural networks (RNN). 

\subsection{Neural Variational Embedding}\label{subsec:ve}
Suppose we have $n$ nodes. We first express each node as a binarized sparse vector with one-hot encoding. Denote the input feature of the $i$th node as $u_i$. For simplicity, we only use the identity of the node as the input feature, i.e. $u_i$ is a binary vector with $i$th entry being $1$ and other entries being $0$. Note that $u_i$ can be easily extended to content-based or neighbor-based features. 

\begin{figure}[h]
\centering
\includegraphics[width=0.60\textwidth]{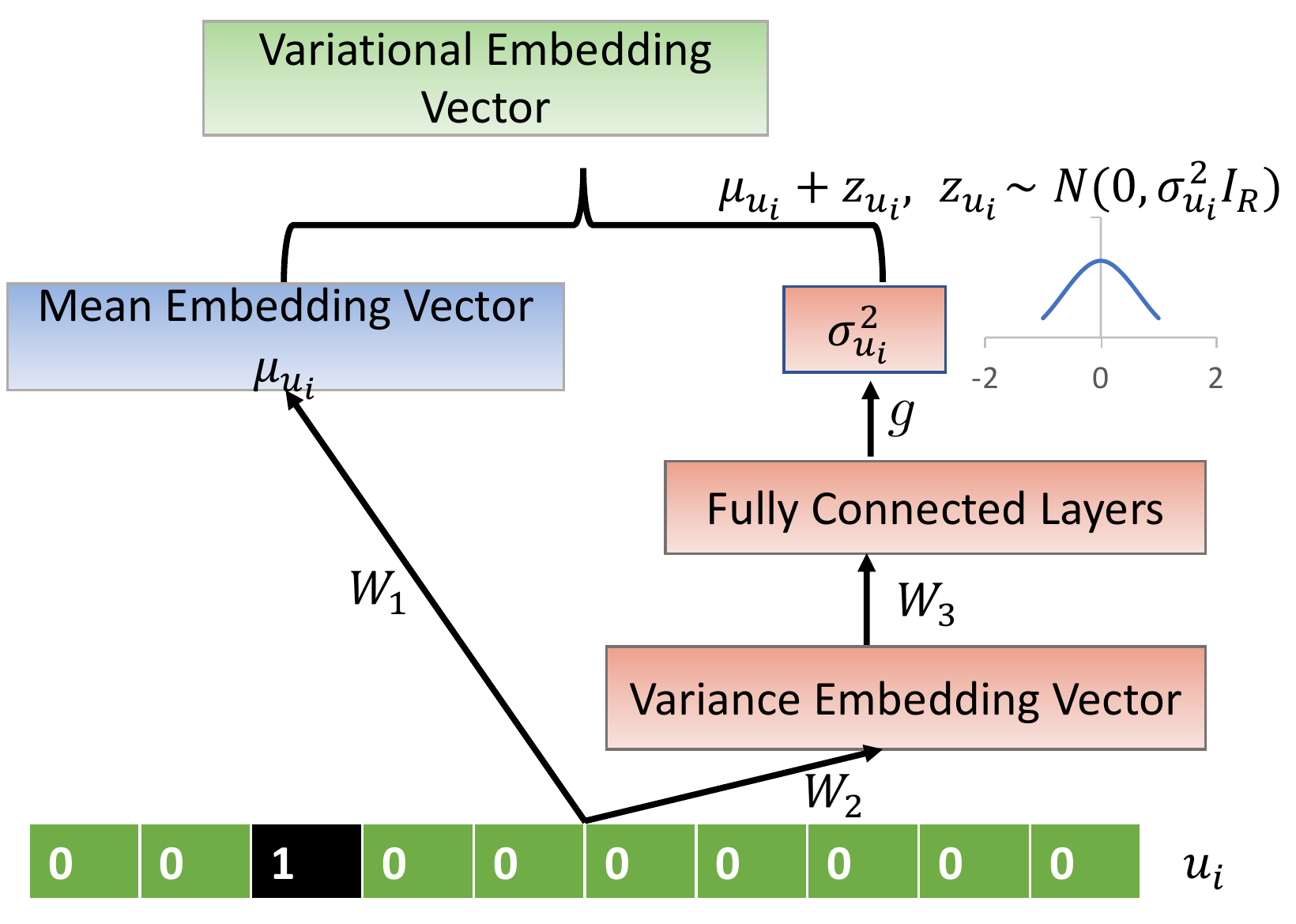}
\caption{Variational embedding architecture.}
\label{fig:VE}
\end{figure}

Denote the embedding feature for each node as $w_i$. Suppose $w_i$ follows the regression function as 
\begin{equation}\label{eq:embed_decom}
w_i = W_1 u_i + z_{u_i},
\end{equation}
where $W_1 \in \mathbb{R}^{R\times n}$ is the mean embedding matrix, $z_{u_i}\sim N(0, \sigma_{u_i}^2 I_R)$ with $I_R$ as an $R\times R$ identity matrix, and $R$ is the embedding dimension. Define $\mu_{u_i}= W_1 u_i$, i.e., the $i$th column of $W_1$. E.q. (\ref{eq:embed_decom}) says that our embedding feature consists of two parts: the mean $\mu_{u_i}$ and the random variation $z_i$ induced by $\sigma_{u_i}$. As shown in Figure \ref{fig:VE}, the mean embedding vector $\mu_{u_i}$ can be achieved via learning $W_1$; and we learn the variance $\sigma_{u_i}^2$ from the variance embedding vector through fully connected layers. That is,
$$
\sigma_{u_i}^2 = g(W_3W_2u_i),
$$
where $W_2$ and $W_3$ are weight matrices. To guarantee $\sigma_{u_i}^2\geq 0$, the activation function $g$ for the output layer can be chosen from the following candidates based on the performance. 
\begin{equation}\label{eq:active_function}
 g(x)=
    \begin{cases}
      \max\{0,x\}, \\ %
      |x|, & \\ %
      x^2. & %
    \end{cases}
\end{equation}
Then following (\ref{eq:embed_decom}), the embedding feature $w_i$ can be achieved by combining $\mu_{u_i}$ 
and $z_{u_i}$ generated from $N(0, \sigma_{u_i}^2I_R)$. 

\subsection{Dynamic Variational Embedding}
In sequence-aware data, the variation changes dynamically, which further leads to the dynamic embedding for each node. Denoting $w_i^{(t)}$ as the embedding feature of node $i$ at time $t$, we update (\ref{eq:embed_decom}) to the following temporal model
\begin{equation}\label{eq:embed_dve}
w_i^{(t)} = W_1 u_i^{(t)} + z_{u_i}^{(t)},
\end{equation}
where $u_i^{(t)}$ is the input feature of the $i$th node at time $t$, and $z_{u_i}^{(t)}$, representing the variational part, is generated from $N(0, \sigma_{u_i}^{2(t)}I_R)$. We incorporate recurrent neural networks (RNN) to learn $\sigma_{u_i}^{2(t)}$, which is different from the variational learning in Section \ref{subsec:ve}.  
RNNs are powerful sequence models that take as their input not just the current input example they see, but also what they have perceived previously in time. However, it is well-known that vanilla RNNs suffer from the vanishing gradient problem. Long short-term memory units (LSTM) are a special kind of RNNs that retains similar structure to the vanilla RNN, but can solve the problem of vanishing and exploding gradients faced while training vanilla RNNs. In this part, we utilize the LSTM to train the dynamic variation of each node. 

\begin{figure}[h]
\centering
\includegraphics[width=0.60\textwidth]{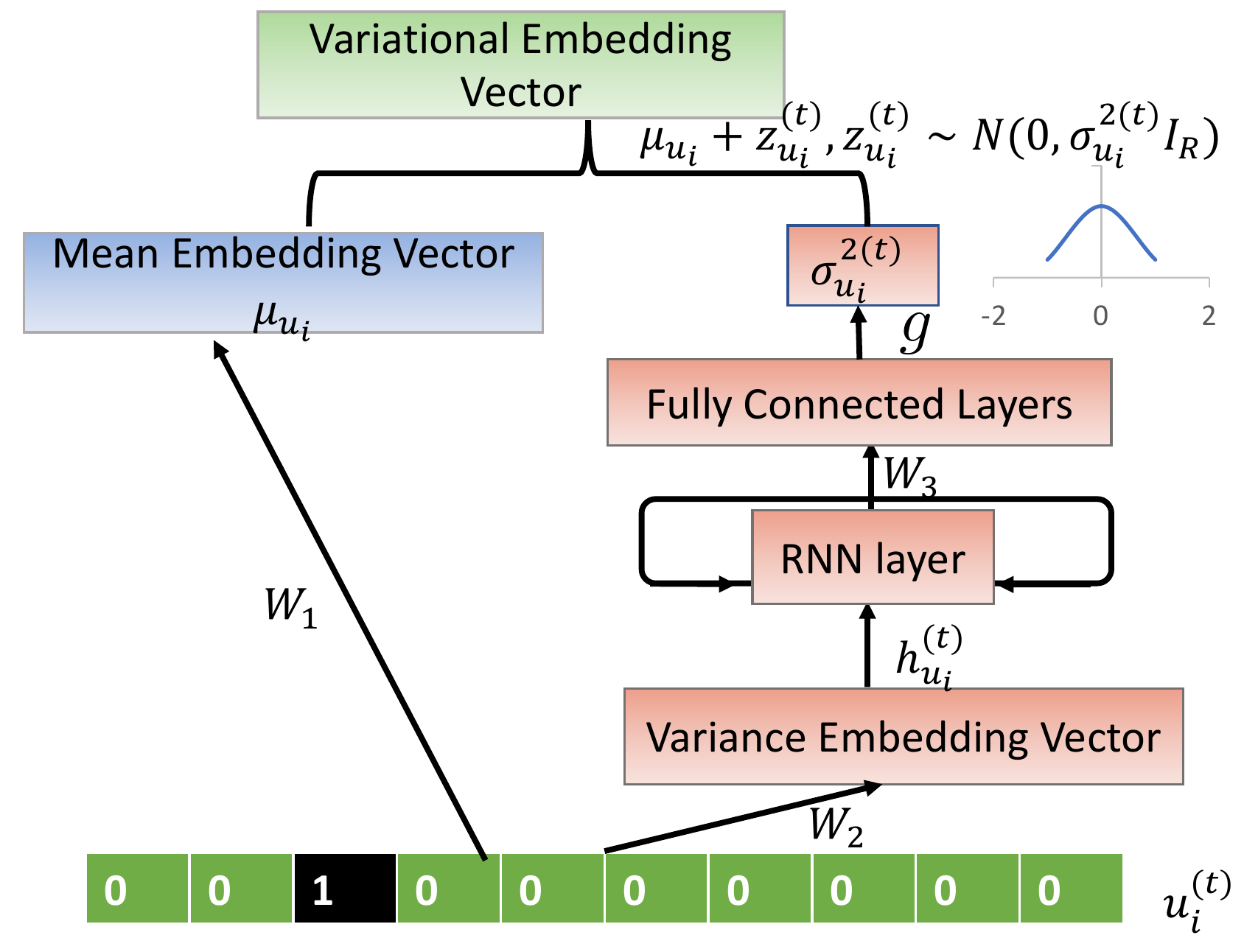}
\caption{Dynamic variational embedding architecture.}
\label{fig:DVE}
\end{figure}

As shown in Figure \ref{fig:DVE}, the variance embedding vector $W_2u_i^{(t)}$ is fed into a recurrent neural architecture. The output dense vector depends on the current history $h_{u_i}^{(t-1)}$ by means of a recurrent layer $h_{u_i}^{(t)}$:
$$
h_{u_i}^{(t)} = RNN(h_{u_i}^{(t-1)}, W_2u_i^{(t)}).
$$
The $h_{u_i}^{(t)}$ is then fed into the fully connected layers via the weight matrix $W_3$, and finally outputs the dynamic variance $\sigma_{u_i}^{2(t)}$ based on the activation function $g$ specified in (\ref{eq:active_function}). 

After obtaining $\sigma_{u_i}^{2(t)}$, we generate $z_{u_i}^{(t)}$ from $N(0, \sigma_{u_i}^{2(t)} I_R)$, where $I_R$ is the identity matrix with dimension $R$. The final DVE of the $i$th node can be achieved by combining $\mu_{u_i}$ and $z_{u_i}^{(t)}$ by e.q. (\ref{eq:embed_dve}).

\section{DVE-based Neural Collaborative Filtering }
In this section, we apply the DVE to sequence-aware recommender systems, and construct a neural collaborative filtering architecture to learn the model parameters and the user-item interaction. We first provide a brief introduction of the graph notation in recommender systems.

\subsection{Notations in Recommender Systems}
Denote $G= (U,V,Y)$, where $U$ consists of $n$ users, $V$ consists of $m$ items, $Y= (Y^{(1)}, \cdots, Y^{(T)})$, and each $Y^{(t)}$ is an $n\times m$ incidence matrix, with each entry $y^{(t)}_{ij}$ denoting the value of the interaction between user $i$ and item $j$ at time $t$, where $i=1, \cdots, n$, $j=1, \cdots, m$, $0<t \leq T$. For example, in e-commercial recommender systems,  $y^{(t)}_{ij} =0,1,2,3$ represents that user $i$ has no access/no response, click, add to cart and buy actions on item $j$ at time $t$, respectively. In recommender rating systems,  $y^{(t)}_{ij}=0,1,\cdots, 5$ denotes the possible ratings of user $i$ on item $j$ at time $t$: $0$ denotes no access, $1$ denotes a poor rating and $5$ is the maximum value allowed. Denote $W^{(t)}= (w^{(t)}_1, \cdots, w^{(t)}_n)\in \mathbb{R}^{R\times n}$ as the embedding matrix of $n$ users at time t, with $w^{(t)}_i\in \mathbb{R}^{R\times 1}$ as the embedding feature of the $i$th user at time $t$. Similarly, define $Q^{(t)} = (q^{(t)}_1, \cdots, q^{(t)}_m)\in \mathbb{R}^{R\times m}$ as the embedding matrix of $m$ items at time $t$, with $q^{(t)}_j\in \mathbb{R}^{R\times 1}$ as the embedding feature of the $j$th item at time $t$. 

Define $u_i^{(t)}$ and $v_j^{(t)}$ as the input feature of user $i$ and item $j$ at time $t$, respectively. In fact, for user $i$, $u_i^{(1)} = \cdots = u_i^{(T)}= u_i$; for item $j$, $v_j^{(1)} = \cdots = v_j^{(T)}= v_j$, with $u_i$, $v_j$ encoded following Section \ref{subsec:ve}. The purpose of introducing the index ${(t)}$ here is to activate the current history in the RNN layer when learning DVE. Notations are summarized in Table \ref{table:notation}. 

\begin{table}[h]
\small
    \centering
\begin{tabular}{c|p{0.3\textwidth}}
\hline\hline
{\bf{Notation}} & {\bf{Description}}\\
\hline
 U & the set of $n$ users\\
 V & the set of $m$ items\\
 $Y^{(t)}$ & the $n\times m$ incidence matrix at time $t$ \\
 $W^{(t)}$ & the embedding matrix of users at $t$\\
 $w_i^{(t)}$ & the embedding feature of user $i$ at $t$\\
 $Q^{(t)}$ & the embedding matrix of items at $t$\\
 $q_j^{(t)}$ & the embedding feature of item $j$ at $t$\\
 $u_i^{(t)}$ & the input feature of user $i$ at $t$\\
 $v_j^{(t)}$ & the input feature of item $j$ at $t$\\
  \hline\hline
\end{tabular}
\caption{Notations.}\label{table:notation}
\end{table}

\subsection{Dynamic Neural collaborative filtering}
\begin{figure}[h]
\centering
\includegraphics[width=0.70\textwidth]{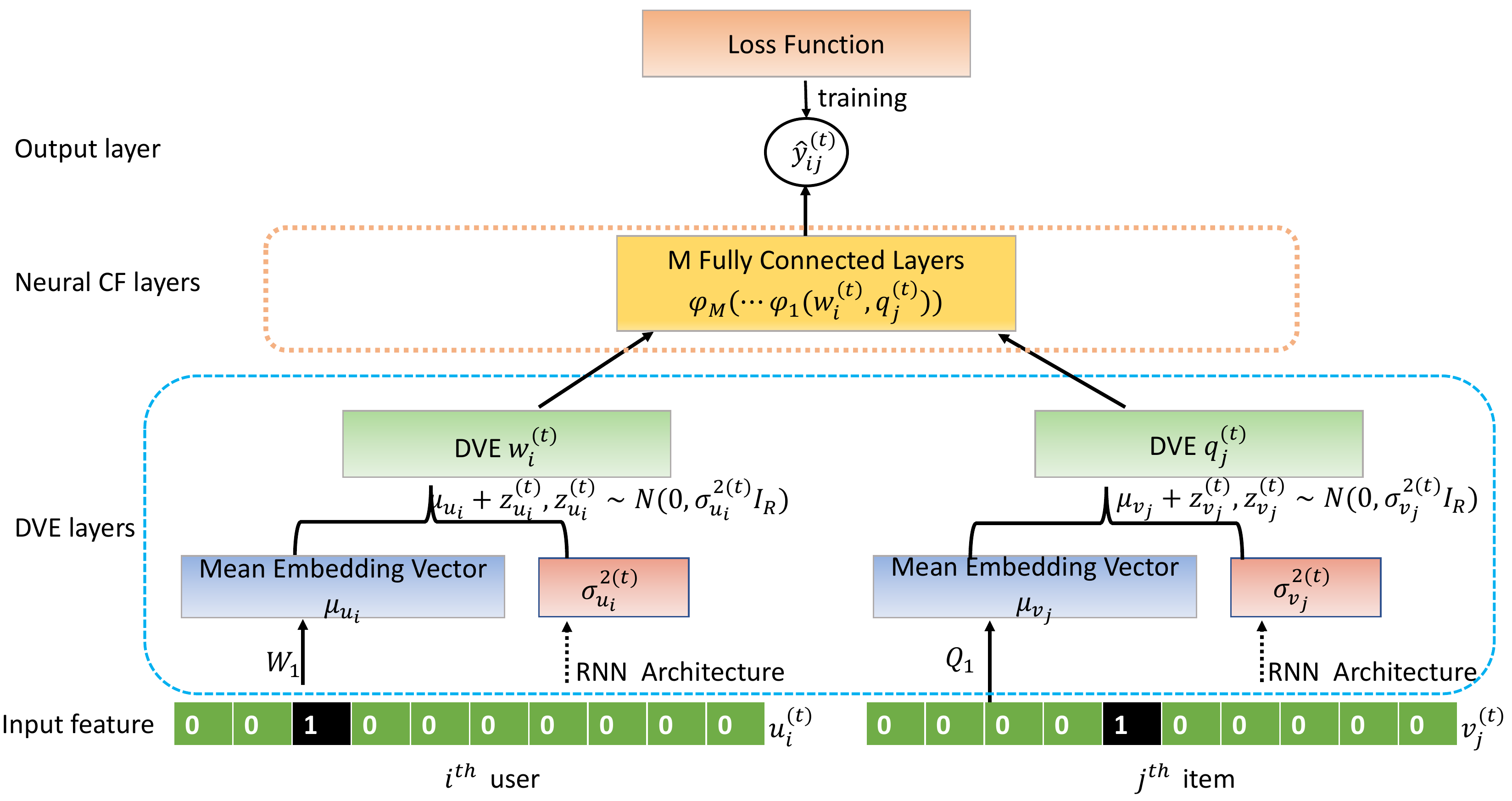}
\caption{Neural collaborative filtering architecture with DVE. }
\label{fig:NCF}
\end{figure}

Collaborative filtering predicts what items a user will prefer by discovering and exploiting the similarity patterns across users and items. Here we use the DVE layers to learn user/item embedding features. Inspired by \cite{he2017neural}, we construct the neural collaborative filtering (NCF) layers, and combine them with the DVE layers to learn the model parameters and the user-item interaction. Figure \ref{fig:NCF} illustrates the dynamic NCF architecture. As shown in Figure \ref{fig:NCF}, we fed the user/item embeddings based on DVE into a multi-layer neural architecture, and finally output the predicted score $\hat{y}_{ij}^{(t)}$. The training is performed by minimizing the loss function as specified in the following part.  

Given the embedding feature matrix $W^{(t)}$ and $Q^{(t)}$, the predicted score between user $i$ and item $j$ at time $t$ can be expressed as 
\begin{equation}\label{eq:est_y}
\hat{y}_{ij}^{(t)} = f(W^{(t)}u_i^{(t)}, Q^{(t)}v_j^{(t)} | W^{(t)}, Q^{(t)}),
\end{equation}
where $f(\cdot)$ is the interaction function defined as 
\begin{equation*}
  f(W^{(t)}u_i^{(t)}, Q^{(t)}v_j^{(t)}) =  \varphi_{out}\Big(\varphi_M\big(\dots \varphi_1(W^{(t)}u_i^{(t)}, Q^{(t)}u_j^{(t)})\big)\Big),
\end{equation*}
where $\varphi_{out}$ and $\varphi_M$, respectively, denote the mapping function for the output layer and the $M$-th neural collaborative filtering layer, and there are $M$ NCF layers in total. Therefore, $M$ determines the model's learning capacity. Note that $w_i^{(t)} = W^{(t)}u_i^{(t)}$ is user $i$'s embedding feature obtained via DVE, i.e., model (\ref{eq:embed_dve}). Similarly, $q_j^{(t)}=Q^{(t)}u_j^{(t)}$ is the item $j$'s embedding feature obtained via DVE with  
$
q_j^{(t)} = Q_1 v_j^{(t)} + z_{v_j}^{(t)}. 
$

Commonly used feedbacks in RS include two categories: explicit (e.g., ratings, votes) and implicit (e.g., clicks, purchases). Explicit feedback data are often in the form of numeric ratings from users to express their preferences regarding specific items. In this case, we can view $y_{ij}^{(t)}$ as a continuous variable. We use squared loss to learn model parameters, and the loss function is defined as 
\begin{equation}
\mathcal{L}_{sq} = \sum_{t=1}^T \sum_{(i,j)\in (U\cup V)^{(t)}} \big(y_{ij}^{(t)} - \hat{y}_{ij}^{(t)}\big)^2,
\end{equation}
where $(U\cup V)^{(t)}$ refers to the observed interaction between user and item nodes at time $t$.

Implicit feedback data are easier to collect, it is also called one-class RS in which only positive implicit feedback can be observed. The target value $y_{ij}^{(t)}$ is $1$ if user $i$ and item $j$ have interaction at time $t$, and $0$ otherwise. For the binary response case, in order to guarantee $\hat{y}_{ij}^{(t)}\in \{0,1\}$, we impose a logistic model on the activation function for the output layer $\varphi_{out}$, i.e., 
$\varphi_{out}(x) =\frac{e^x}{1+e^x}$. Denote $\mathcal{Y}^{(t)}$ as the set of observed interactions in $Y$ at time $t$, and $\mathcal{Y}^{(t)-}$ as the set of negative instances, which can be no interactions or unobserved interactions. Define $\mathcal{Y} = (\mathcal{Y}^{(1)}, \cdots, \mathcal{Y}^{(T)})$, $\mathcal{Y}^{-} = (\mathcal{Y}^{(1)-}, \cdots, \mathcal{Y}^{(T)-})$, $W= (W^{(1)}, \cdots, W^{(T)})$, $Q = (Q^{(1)}, \cdots, Q^{(T)})$, also denote $\Theta$ as the model parameters in the neural architecture. Then the likelihood function can be written as 
$$
P(\mathcal{Y}, \mathcal{Y}^{-}|W, Q, \Theta) =  \prod_{t=1}^T \prod_{(i,j)\in \mathcal{Y}^{(t)}}\hat{y}_{ij}^{(t)} \prod_{(i,j)\in \mathcal{Y}^{(t)-}} (1-\hat{y}_{ij}^{(t)}),
$$
where $\hat{y}_{ij}^{(t)}$ is estimated by e.q.(\ref{eq:est_y}). The log-likelihood loss function can be written as 
$$
\mathcal{L} = \sum_{t=1}^T \sum_{(i,j) \in \mathcal{Y}^{(t)}\cap \mathcal{Y}^{(t)-}} \big(y_{ij}^{(t)} \log \hat{y}_{ij}^{(t)} + (1-y_{ij}^{(t)})\log(1-\hat{y}_{ij}^{(t)})\big). 
$$
In practice, when the interactions are sparse, we uniformly sample from the negative sets $\mathcal{Y}^{(t)-}$ at each time $t$, and control the sampling ratio to the range between $1:3$ and $1:5$.

\section{Experimental Results}
In this section, we implement our proposed DVE method on sequence-aware recommender systems, and compare with existing methods using one of the most popular public data sets: Movielens. We first compare our proposed method with several competitors which are designed for explicit data. We further exam its performance on implicit data, by transforming the ratings into 0 or 1 based on whether the user has rated the item or not. 
The numerical studies are run on a computing workstation with two Titan-V GPU processors and 64GB RAM.

\subsection{Movielens data description}
The Movielens-1M data set is collected by GroupLens Research and is downloaded from \url{http://grouplens.org/datasets/movielens}. It contains $1,000,209$ ratings of $3883$ movies by $6040$ users, and the rating scores range from 1 to 5. The data are collected from April 2000 to February 2003. The timestamps are recorded to show when a user rates a movie. 

We observe that the variation of the popularity of the movie and the preference of the user dramatically changes over time. For example, the number of viewers might be large in the first few months following its release date, then decrease after that. We plot the number of ratings versus time for the movie titled as ``The Perfect Storm'' in Figure \ref{fig:movielens}(a). It shows that the number of ratings is increasing from May to December of $2000$ and dropped to less than $20$ in $2001$. The strong temporal pattern motivates us to model the variance dynamically. Movies having features with large variance will tend to be recommended to a broader range of users, while lower feature variance narrows the recommendation range. 

\subsection{MovieLens data with explicit feedback} 
We directly use the rating as explicit feedback. The proposed method is compared with the following competitors designed for explicit feedback recently. 
\begin{itemize}
\item \citet{agarwal2009regression} proposes a regression-based latent factor model.
\item \citet{mazumder2010spectral} provides a soft-impute algorithm to replace the missing elements with those obtained from a soft-thresholded SVD.
\item \citet{zhu2016personalized} proposes a likelihood method to seek a sparse latent factorization, from a class of overcomplete factorizations, possibly with a high percentage of missing values. 
\item  \citet{he2017neural} establishes the general NCF framework based on one-hot embedding layer for latent features of each user and item. 
\item \citet{bi2017} proposes a group-specific method to use dependency information from users and items which share similar characteristics under the singular value decomposition framework.
\end{itemize}

We first order all the ratings based on their timestamps. Then we set the first $75\%$ as the training data set and set aside the last $25\%$ of ratings as the testing data set.
The root mean square error (RMSE) of the testing set is reported in Table \ref{tab:1}.

\begin{table}[h!]
\centering
\begin{tabular}{l|c}
  \hline
\multicolumn{2}{c}{Dataset: Movielens-1m}\\
\hline
 Method & RMSE\\
\hline
DVE (our proposed) & \textbf{0.891} \\
 \citet{agarwal2009regression} & 1.197  \\
 \citet{mazumder2010spectral} & 1.073 \\
 \citet{zhu2016personalized}& 1.063\\
 \citet{bi2017} & 0.964 \\
 \citet{he2017neural} & 0.933 \\ 
 \hline
\end{tabular}
\caption{Experimental results of Movielens-1M dataset with explicit feedback. }\label{tab:1}
\end{table}

Table \ref{tab:1} provides the prediction results on the testing set, and shows that our proposed method outperforms other methods significantly. The RMSE of the proposed method is 25.6\% less than \citet{agarwal2009regression}, 17.0\% less than \citet{mazumder2010spectral}, 16.2\% less than \citet{zhu2016personalized}, 7.6\% less than \cite{bi2017},
and 4.5\% less than \citet{he2017neural}.

\subsection{Movielens data with implicit feedback} 
In this setting, we code the user-movie interaction as a binary variable in which $1$ indicates that the user rates the movie and $0$ indicates that the rating is missing. %

We use leave-one-out evaluation to evaluate the performance of item recommendation; see \cite{he2017neural,bayer2017generic,he2016fast} etc. For each user, we hold-out his/her latest interaction as the test item and utilize the remaining data for training. To increase the computational efficiency, we randomly sample $100$ items that are not interacted with by the user, and rank the test item among the $100$ items. 

The evaluation is done on top-$k$ recommendation. The performance of a ranked list is judged by the overall {\it{top-$k$ Hit Ratio}} (HR@k) 
and overall {\it{top-$k$ Normalized Discounted Cumulative Gain}} (NDCG@k). The HR@k measures whether the test item is included in the top-$k$ list. 
NDCG@k gives more weight to the relevant items on top of the recommender list, and is defined as 
\begin{equation}
    NDCG@k = \frac{\sum_{i=1}^k r_i/\log_2(i+1)}{\sum_{i=1}^{k}1/\log_2(i+1)},
\end{equation}

The performance of a ranked list is judged by the averaged HR@10 and NDCG@10 for all users. We compare with \cite{he2017neural} for their three methods: the generalized matrix factorization method (GM), the pure neural network framework with one-hot embeddings (MPL), and a fusion of the two (Neural-GM). We implement their algorithm using the docker image provided in \url{https://github.com/hexiangnan/neural_collaborative_filtering}.

\begin{table}[h!]
\centering
\begin{tabular}{l|c|c}
  \hline
\multicolumn{3}{c}{Movielens-1M with implicit feedback}\\
\hline
 Method & NDCG@10 & HR@10 \\
\hline
Our proposed & \textbf{0.4211} & \textbf{0.6924}\\
 \citet{he2017neural}-GM & 0.3676 & 0.6358 \\
 \citet{he2017neural}-MPL & 0.3942 & 0.6737 \\
 \citet{he2017neural}-Neural-GM & 0.4073 & 0.6790 \\
 \citet{liang2018variational}-VAE& 0.0416 & 0.4481\\
 \hline
\end{tabular}
\caption{Experimental results for Movielens-1M data set with implicit feedback.}\label{tab:s2}
\end{table}

 Table \ref{tab:s2} shows that our proposed method outperforms the state-of-the-art methods \citet{he2017neural}-GM, \citet{he2017neural}-MPL and \citet{he2017neural}-Neural-GM by a large
3\% -15\% and 2\%-9\%, respectively. We also compare our proposed method with the VAE-based collaborative filtering studied in \cite{liang2018variational}. As shown in Table \ref{tab:s2}, the NDCG@10 is lower than $0.1$, and HR@10 is lower than $0.5$. In \cite{liang2018variational}, they use the evaluation strategy by holding out several users as the testing set. However, in this study, for each user, we hold out the latest interaction as the testing set for evaluation. The VAE-based approach has limitation in exploration under such evaluation strategy, since its input treats all the unobserved movies as $0$ and the VAE is designed to learn the representation of the input. Instead, our proposed method uses negative sampling to sample a small portion of the negative set, thus it turns out to have better performance in exploration for existing users. 

\section{Discussion}
We propose a dynamic variational embedding framework and implement it for collaborative filtering with temporal information. Our method is simple and generic; it is not limited to the applications presented in this paper, but is designed to any embedding task. This work complements the mainstream embedding models by incorporating variation and dynamic changes, opening up a new avenue of research possibilities for wide range of embedding models. In the future, we will study the knowledge-based embedding to model auxiliary information, such as user reviews, user geographical information, and movie reviews. Individualized or itemized information could help us to better understand the uncertainty and dynamic pattern of the embeddings.

\bibliographystyle{plainnat}
\bibliography{ijcai19}

\end{document}